\documentclass[12pt]{iopart}
\usepackage{graphicx}
\usepackage{iopams}  
\begin{document}

\newcommand{\bea}{\begin{eqnarray}}
\newcommand{\eea}{\end{eqnarray}}
\newcommand{\be}{\begin{equation}}
\newcommand{\ee}{\end{equation}}
\newcommand{\bse}{\begin{subequations}}
\newcommand{\ese}{\end{subequations}}
\newcommand{\p}{\partial}
\renewcommand{\ss}[1]{_{\hbox{\tiny #1}}}

\paper[]{Dynamics versus energetics in phase separation}

\author{Paolo Politi and Alessandro Torcini}

\address{CNR - Consiglio Nazionale delle Ricerche - Istituto dei Sistemi Complessi, 
Via Madonna del Piano 10, 50019 Sesto Fiorentino, Italy}
\ead{Paolo.Politi@cnr.it, Alessandro.Torcini@cnr.it}

\begin{abstract}
Phase separation may be driven by the minimization of a suitable free energy ${\cal F}$. 
This is the case, e.g., for diblock copolimer melts, where ${\cal F}$ is minimized by a steady periodic pattern
whose wavelength $\lambda\ss{GS}$ depends on the segregation strength $\alpha^{-1}$ and 
it is know since long time that in one spatial dimension $\lambda\ss{GS} \simeq \alpha^{-1/3}$.
Here we study in details the dynamics of the system in 1D for different initial conditions 
and by varying $\alpha$ by five orders of magnitude.
We find that, depending on the initial state, the final configuration
may have a wavelength $\lambda\ss{D}$ with
$\lambda\ss{min}(\alpha)<\lambda\ss{D}<\lambda\ss{max}(\alpha)$, where $\lambda\ss{min} \approx
\ln (1/\alpha)$ and $\lambda\ss{max}\approx \alpha^{-1/2}$.  
In particular, if the initial state is homogeneous, the system exhibits 
a logarithmic coarsening process which arrests whenever $\lambda\ss{D}\approx\lambda\ss{min}$.
\end{abstract}

\vspace{2pc}

%
%
%

\section{Introduction}

A physical system quenched from a disordered, homogeneous phase into a coexistence region
undergoes a phase ordering process~\cite{Onuki} which is characterized by the minimization of the free energy ${\cal F}$.
Very similar dynamics may appear in driven out of equilibrium systems~\cite{Cross_Greenside}, 
where the free energy is
replaced by a suitable Lyapunov functional ${\cal F}\ss{NE}$.
In both cases the final, equilibrium (or steady) state is expected to be the state which minimizes
${\cal F}$ (or ${\cal F}\ss{NE}$). Often, this minimal energy configuration is characterized by a finite
wavelength or it corresponds to a complete phase separation, achieved via a coarsening process.
As a matter of fact, the former picture (final fixed periodicity) can also be attained via a
transient coarsening process.

It is worth to mention two limiting cases, the ordering of a binary alloy and the Rayleigh-B\'enard
pattern formation close to the convective threshold. The former problem can be described by the Cahn-Hilliard
(CH) equation~\cite{Bray}, which is known to be characterized by a perpetual coarsening process.
The second phenomenon is usually described by the Swift-Hohenberg (SH) equation~\cite{Cross_Greenside}, 
whose behavior close to the
instability threshold is universal for stationary type-I instabilities~\cite{Cross_Greenside}:
it is characterized by the emergence of a periodic pattern 
whose modulation 
undergoes a secondary instability, the so-called Eckhaus instability~\cite{Collet_Eckmann},
a transient process during which rolls are locally created or destroyed until a stable roll size is attained.

Both CH and SH equations are characterized by an homogeneous phase which is linearly unstable
in a range $(q_1,q_2)$ of wavevectors, with the difference that $q_1=0$ for CH and $q_1\simeq q_2 \simeq q_c$ for SH.
However, there is an even more important difference between the two. All CH periodic steady states are unstable,
while SH periodic steady states are stable in a finite $q-$interval around the value 
which minimizes the
SH free energy. The existence of such stable interval cannot ensure the ground state is dynamically attained.
In the absence of noise it is almost certainly not attained and even if noise is present, 
the time scale might be
too large to be experimentally observed. However, in the SH problem near the threshold, the final 
state is always very close to the ground state, because the interval $(q_1,q_2)$ is a small interval around a finite value $q\ss{c}$.

A physical problem which has long been studied~\cite{Oono_Shiwa,Liu_Goldenfeld,bates1990,PhysicsToday,vgoono_co,vgoono_en,Benilov,vgoono_pd} 
and which is related to both limiting cases discussed here above (CH and SH) 
is the phase separation in diblock copolymer melts. 
In this example two different subchains $A$ and $B$ compose a chain molecule. 
In the disordered phase $A$ and $B$ are well mixed, but below $T\ss{c}$ there is a tendency to segregation
which is never complete, because of a long range term in the free energy which stabilizes the system
at a certain length scale depending on the strength of segregation, $\alpha^{-1}$ (see next Section). 
For infinite segregation strength $(\alpha=0)$
the system behaves as a binary alloy and it is described by the CH equation. For weak segregation there
is a rapid selection of the size of the pattern, described by the SH equation.

In the seminal paper by Liu and Goldenfeld~\cite{Liu_Goldenfeld}, authors focus on
$\lambda\ss{GS}$, the wavelength of the pattern which minimizes the free energy.
They find numerically $\lambda\ss{GS}\approx \alpha^{-1/4}$ for $\alpha>10^{-2}$ and
$\lambda\ss{GS}\approx \alpha^{-1/3}$ for vanishing $\alpha$.
We are interested to the second regime and all our results refer to the limit of very small $\alpha$.
The other regime has been recently addressed by Benilov et al.~\cite{Benilov}, who study in detail how the
region of stable patterns widens with decreasing $\alpha$, starting from the SH limit.

Comprehensive studies on the dynamics of this system with varying $\alpha$
and with varying initial conditions are absent. Therefore,
in this manuscript we focus on dynamics in the regime of small $\alpha$, showing the existence of a range 
$(q\ss{s1},q\ss{s2})=(2\pi/\lambda\ss{max},2\pi/\lambda\ss{min})$ 
of wavevectors where steady states are stable and whose extrema both vanish with $\alpha$, according to well
defined laws: $q\ss{s1}\approx\sqrt{\alpha}$ and $q\ss{s2}\approx -1/\ln\alpha$. 
If phase ordering starts from the disordered state, the final configuration is 
essentially determined by $q\ss{s2}=2\pi/\lambda\ss{min}$, which is attained via a logarithmic
coarsening process.

In Sec.~\ref{model} we define the model, set the relevant notations and questions,
and discuss the energetics.
In Sec.~\ref{dynamics} we report our results on dynamics, which are discussed in the final Sec.~\ref{discussion}.

\section{The model and the energetics}
\label{model}

The model equation we are using to study the problem of phase separation is the so-called Oono-Shiwa~\cite{Oono_Shiwa} 
(OS) equation,
\be
\frac{\p u}{\p t} = -\p_{xx}\left( u_{xx} + u-u^3\right) -\alpha u
\label{oono}
\ee
which depends on one parameter only, $\alpha$, because the coefficients of the other terms 
can be set to one by a suitable rescaling of $u,x,$ and $t$.

\begin{figure}
\begin{center}
\includegraphics[width=0.49\columnwidth]{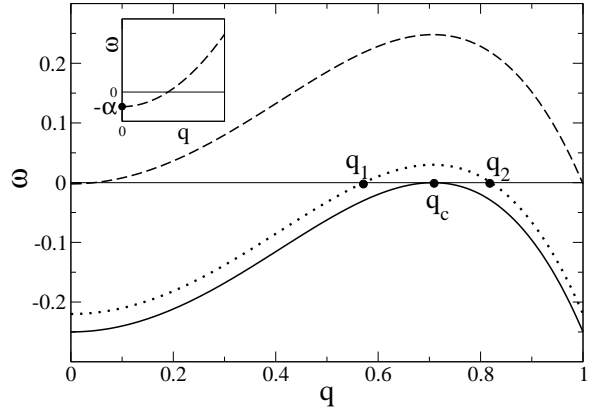}
\end{center}
\caption{
The linear stability spectrum for $\alpha=0.002,0.22,0.25$ (from top to bottom).
The explicit values of $q_{1,2}(\alpha)$ are given in Eq.~(\ref{q12}).
The case $\alpha=0.002$ (dashed line) is enlarged in the inset around $q=0$,
to stress that $\omega(q\to 0)<0$ for any $\alpha>0$.
}
\label{fig_stability}
\end{figure}

The linear stability analysis of the homogeneous phase $u\equiv 0$ is easily obtained assuming
$u(x,t)=u_0\exp(\omega t +iqx)$ and linearizing Eq.~(\ref{oono}) in $u_0$. We get
\be
\omega(q)=-\alpha +q^2 - q^4,
\label{eq_omega}
\ee
which is plotted in Fig.~\ref{fig_stability}. Therefore, the equation is linearly unstable $(\omega(q)>0$) 
for $q_1<|q|<q_2$, where
\be
q_1 = \left( \frac{1}{2} - \sqrt{\frac{1}{4} -\alpha}\, \right)^{1/2} \quad \mbox{and} \quad
q_2 = \left( \frac{1}{2} + \sqrt{\frac{1}{4} -\alpha}\, \right)^{1/2}
\label{q12}
\ee
and the most unstable mode is $q\ss{c}=\frac{1}{\sqrt{2}}$, with $\omega(q\ss{c})= \frac{1}{4}-\alpha$.

The Oono-Shiwa equation trivially reduces to the Cahn-Hilliard equation for $\alpha=0$. 
It is well known~\cite{Langer1971} that in this limit 
we get perpetual logarithmic coarsening and the asymptotic profile is a sequence of kinks (and antikinks)~\cite{Kawakatsu1985},
which represent finite size domain walls interpolating between the stable solutions $u=-1$ and $u=+1$ 
(or viceversa). However, the OS equation has another interesting limit: for $\alpha\to(\frac{1}{4})^-$,
it displays the well known Eckhaus instability.
We expect this limit from Eq.~(\ref{eq_omega}) and from the nature of the nonlinear term, which is a cubic term.
In fact, in the limit $\alpha\to(\frac{1}{4})^-$ the range of stable states has been determined in~\cite{Benilov}
and it coincides with the Eckhaus scenario, which is reproduced,
for the sake of completeness, in~\ref{app_Eckhaus}.

More precisely, the existence of this limit means that,
if $\alpha=\frac{1}{4}-\epsilon^2$ with $\epsilon\ll 1$, the linearly unstable
interval of the homogeneous phase is $(q_1,q_2)=(q_c-\delta,q_c+\delta)$, with $\delta=\epsilon/\sqrt{2}$.
For any $q$ belonging to such interval there is a steady state, but only those belonging to the narrower
subinterval $(q_c-\delta/\sqrt{3},q_c+\delta/\sqrt{3})$ are stable. 

Let us now introduce the free energy whose minimization drives the dynamics.
The OS equation can be written in a variational form~\cite{Oono_Shiwa},
\be
\p_t u(x,t) = \p_{xx} \frac{\delta {\cal F}\ss{OS}}{\delta u}
\ee
where
\be
{\cal F}\ss{OS}[u] = \int dx \left[ \frac{1}{2} u_x^2 + \left( -\frac{u^2}{2} + \frac{u^4}{4}\right)\right]
- \frac{\alpha}{4} \int dx \int dx' u(x,t)|x-x'|u(x',t).
\qquad
\ee
If we define the density of free energy, $\tilde {\cal F}\ss{OS}[u]= L^{-1}{\cal F}\ss{OS}[u]$ with $L$ meaning the
total extension of the system in the $x-$direction, for a periodic configuration of wavelength $\lambda$ we get
\be
\tilde {\cal F}\ss{OS}[u] =
\left\langle\left[ \frac{1}{2} u_x^2 + \left( -\frac{u^2}{2} + \frac{u^4}{4}\right)\right] \right\rangle
- \frac{\alpha}{4\lambda} \int_{-\lambda/2}^{\lambda/2} dx \int_{-\lambda/2}^{\lambda/2} dx'
u(x) |x-x'| u(x') . \qquad
\label{Fos}
\ee

The former expression cannot be evaluated analytically in an exact manner, because steady states of
OS Eq.~(\ref{oono}) are not known.
However, it has been evaluated either numerically~\cite{Liu_Goldenfeld} or analytically in an
approximate way~\cite{vgoono_en}, giving consistent results: for small $\alpha$, 
$\tilde {\cal F}\ss{OS}[u]$ is minimised
for $\lambda=\lambda\ss{GS}\approx \alpha^{-1/3}$. Because of its simplicity
we give here a brief derivation of this result.

The underlying idea, suggested by numerical sumulations and by dimensional analysis of Eq.~(\ref{oono}),
is the following: if we start from the homogeneous solution, represented by white noise $u(x,t)=\eta(x,t)$,
the term $\alpha u$ which distinguish OS from CH equation comes into play for $t\approx\alpha^{-1}$:
at short times the dynamics is the coarsening process as given by the CH equation, but this process stops when
$t\approx\alpha^{-1}$. If this picture is correct, it is reasonable to evaluate Eq.~(\ref{Fos}) using the steady
states for $\alpha=0$.

The average value $\langle [ \dots ] \rangle$ on the RHS of Eq.~(\ref{Fos}) has the form
$\langle \frac{1}{2} u_x^2 + V(u)\rangle$, with
$V(u)=-(u^2/2)+ (u^4/4)$, and steady states satisfy the equation~\footnote{The
integration of $\partial_t u=0$ for $\alpha=0$ and $\langle u\rangle=0$ gives
$\frac{1}{2}u_x^2 -V(u)=$const, where const$=\frac{1}{4}$ for a kink profile.}
$\frac{1}{2} u_x^2 - V(u) =\frac{1}{4}$, so that
\be
\left\langle \frac{1}{2} u_x^2 + V(u)\right\rangle = \langle u_x^2\rangle - \frac{1}{4}.
\ee
The function $u_x^2$ is concentrated at kink positions. If $\tilde u(x)=\tanh(x/\sqrt{2})$ is the
kink profile centred in $x=0$, we can write
\be
\langle u_x^2\rangle \simeq \frac{4}{\lambda}\int_0^{\lambda/4} dx \frac{1}{2\cosh^4(x/\sqrt{2})} \simeq
\frac{8\sqrt{2}}{\lambda}\left( 1 - e^{-\lambda/\sqrt{2}}\right).
\label{ux2}
\ee
The leading term, $\langle u_x^2\rangle \simeq \frac{8\sqrt{2}}{\lambda}$, is sufficient here.

The term proportional to $\alpha$ in Eq.~(\ref{Fos}) can be evaluated in the simplest way by approximating
the kink profile by a step function, $\tilde u(x)=\pm 1$ for $x\gtrless 0$, getting (see Eq.~(\ref{Fos}))
\be
-\frac{\alpha}{4\lambda} \int_{-\lambda/2}^{\lambda/2} dx \int_{-\lambda/2}^{\lambda/2} dx'
u(x) |x-x'| u(x') \approx \frac{\alpha}{24}\lambda^2.
\ee
Summing up terms in Eq.~(\ref{Fos}) we have
\be
\tilde {\cal F}\ss{OS}[u] =
\frac{8\sqrt{2}}{\lambda} + \frac{\alpha}{24}\lambda^2 - \frac{1}{4} ,
\ee
which has a minimum for
\be
\lambda\ss{GS} =\left( \frac{96\sqrt{2}}{\alpha}\right)^{1/3} \approx \frac{5}{\alpha^{1/3}}.
\ee
Above result, which can be refined within the same spirit~\cite{vgoono_en},
is in agreement with the original numerical analysis by Liu and Goldenfeld~\cite{Liu_Goldenfeld},
which also gives $\lambda\ss{GS} \approx\alpha^{-1/3}$ for small $\alpha$.

\section{Dynamics}
\label{dynamics}

In the previous Section we have shown that, according to energetics, the final wavelength should scale with
the segregation strength as $\alpha^{-1/3}$.
Here we are going to study dynamics instead. For this purpose, 
we have solved numerically the OS equation (\ref{oono}) for
$5 \times 10^{-7} < \alpha < 8 \times 10^{-2}$, using a time splitting pseudo spectral code, whose details are given in~\ref{app_code}.
\begin{figure}
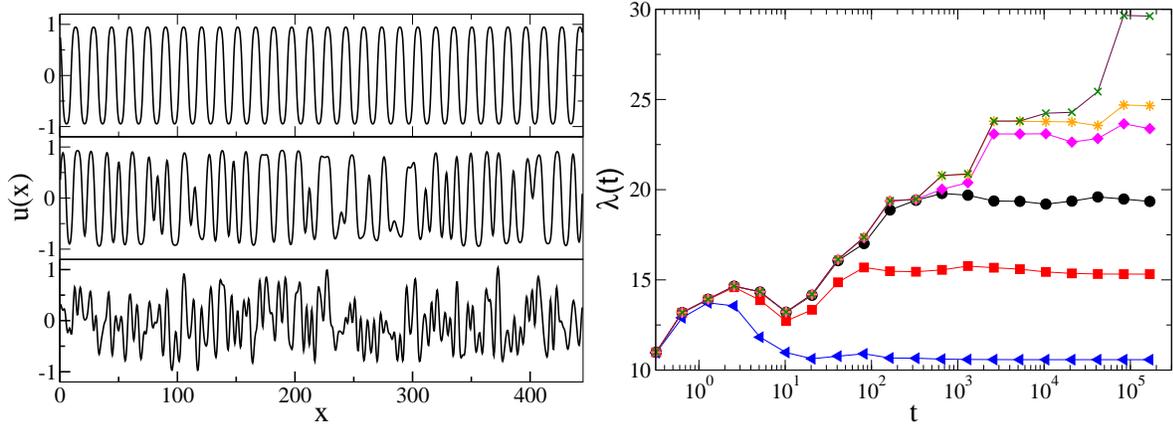

\begin{center}
\includegraphics[width=0.49\columnwidth]{fig2a.eps}
\includegraphics[width=0.49\columnwidth]{fig2b.eps}
\end{center}
\caption{
(Color online)
(a)~Spatial profiles of $u(x,t)$ for $\alpha=10^{-2}$ and $t=0.32,20.5,16777$ (from bottom to top). (b)~The time dependence of the typical wavelength $\lambda(t)$ 
for $\alpha=10^{-1}, 10^{-2}, 10^{-3}, 10^{-4}, 10^{-5}, 10^{-6}$ (from bottom to top).
The data refer to random initial conditions, 
$L=\sqrt{2} \pi \times 100$, time step $\Delta t = 0.0025$ and 1024 Fourier modes
(corresponding to a spatial resolution $\Delta x = 0.43$).}
\label{fig_time}
\end{figure}

In Fig.~\ref{fig_time}(a) we plot the profile for $\alpha=10^{-2}$
at short time (bottom), intermediate time (center), and after
saturation of the wavelength (top). 
Initial conditions correspond to a random profile (more
details are given in~\ref{sec_incon}).
In plate (b) of the same figure
we show the typical wavelength of the structure, $\lambda(t)$, as a function
of time, when varying $\alpha$ (which decreases from bottom to top).
These results give a qualitative proof of a few relevant features
of our system, already discussed in Sec.~\ref{model}: (i)~the system coarsens for a finite time;
(ii)~the coarsening dynamics does not depend on $\alpha$;
(iii)~the final wavelength increases with decreasing $\alpha$.
The non monotonous behavior of $\lambda(t)$ is a well known feature which goes
beyond the OS model (see, e.g., Figs.~7,8 of~\cite{Nicoli} and Fig.~6 of~\cite{torcini2002}).
It has the same interpretation as the non monotonous behavior of the 
roughness~\cite{Castellano_Krug}.

\begin{figure}
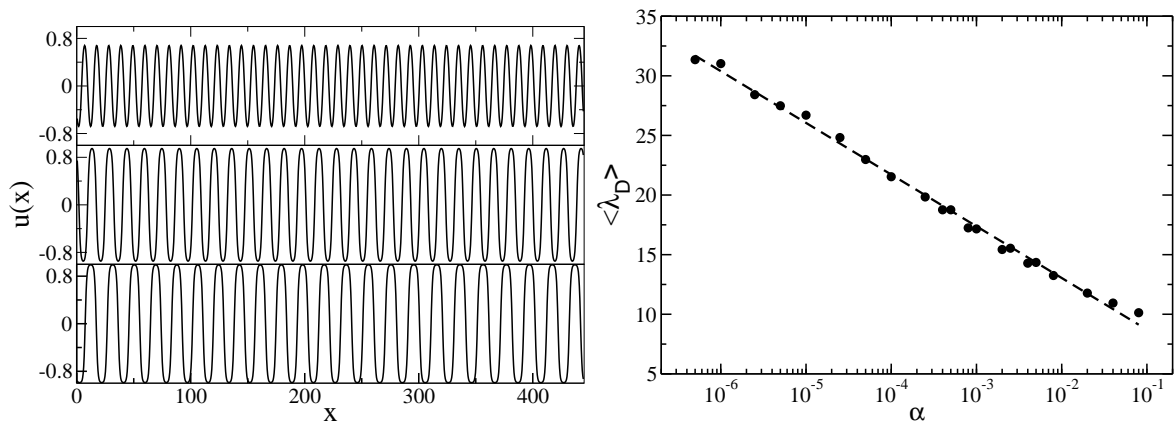

\begin{center}
\includegraphics[width=0.49\columnwidth]{fig3a.eps}
\includegraphics[width=0.49\columnwidth]{fig3b.eps}
\end{center}
\caption{
(a)~Steady spatial profiles $u(x,t\to\infty)$ for 
$\alpha=10^{-1}, 10^{-2}, 10^{-3}$ (from bottom to
top). Shown configurations refer to a time $t= 16777$,
$L=\sqrt{2} \pi \times 100$, time step $\Delta t = 0.0025$ and 1024 Fourier modes.
(b)~The asymptotic length of the structure, $\langle\lambda\ss{D}(\alpha)\rangle=
\langle\lambda(t\to\infty,\alpha)\rangle$, has been obtained by integrating the system up to
time $\simeq 10^5$, starting from a random initial configuration, and by averaging
over $10-30$ different initial conditions (noise realizations). 
The data refer to $L=\sqrt{2} \pi \times 40$, time step 
$\Delta t = 0.0025$ and 1024 Fourier modes, corresponding to $\Delta x = 0.17$.
The straight line is a best fitting and corresponds to the function
$\lambda\ss{D}=4.37 -1.88\ln\alpha$.
}
\label{fig_alpha}
\end{figure}

\begin{figure}
\begin{center}
\includegraphics[width=\columnwidth]{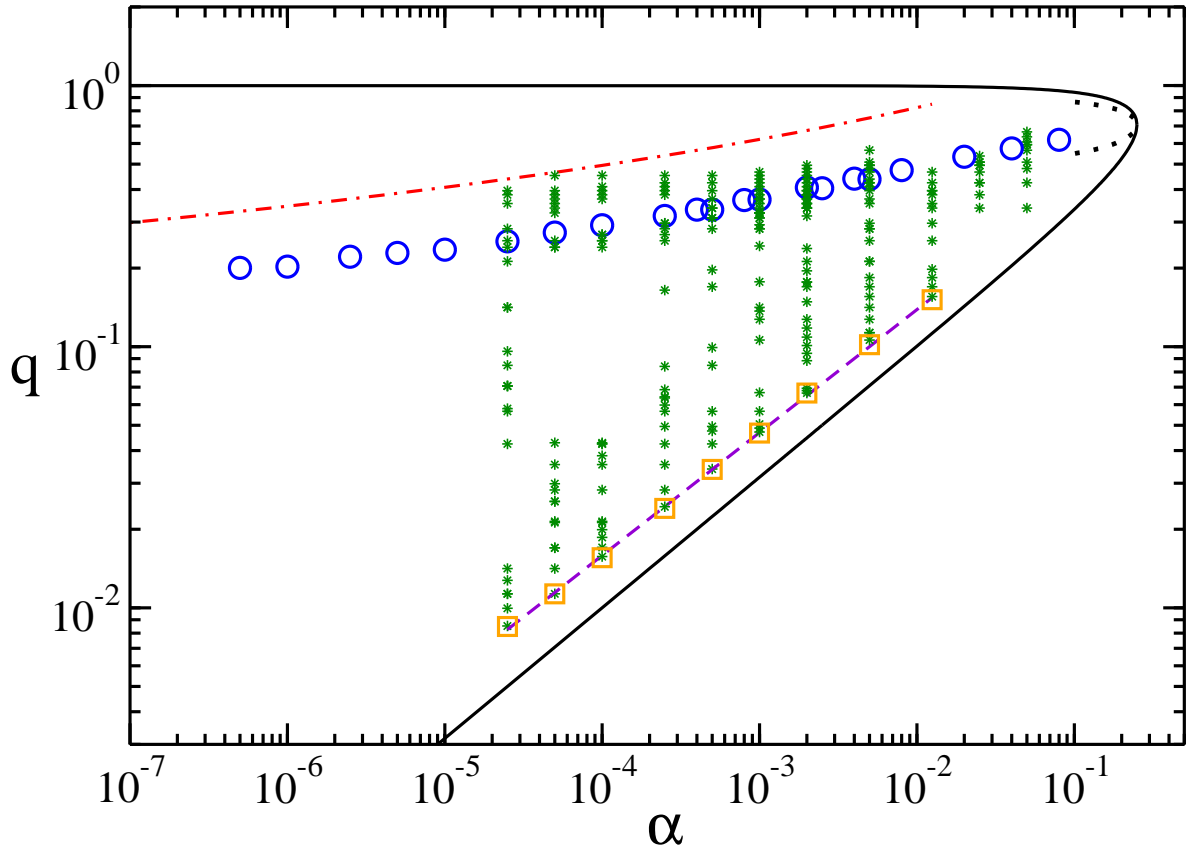}
\end{center}
\caption{
(Color online)
The stability diagram $(q,\alpha)$: the (green) asterisks and
the (blue) circles indicate 
stable solutions found via direct numerical simulations. 
Circles have been obtained by
starting from random initial conditions and they
are the same data plotted in lin-log scale in Fig.~3b.
The initial conditions for asterisks are of various types, 
discussed in the main text and in~\ref{sec_incon}.
(Orange) squares are the minimal stable $q'$s for any $\alpha\le 0.01$.
We also plot the
power law fitting $ \simeq 1.22 \times \alpha^{0.47}$ (dashed violet
line).
The (red) dot-dashed line denotes the theoretical estimate, Eq.~(\ref{theo_estimate}),
of the upper bound for the stability of the wave vectors.
For the details on the performed simulations see~\ref{sec_numerics}. 
Black, dotted lines are the analytical curves
$q\ss{s1,s2}(\alpha)$ in the Eckhaus limit $\alpha\to 1/4$.
}
\label{fig_diagram}
\end{figure}

Now, let us focus on the final configurations, see Fig.~\ref{fig_alpha}.
In (a) we give the profiles at different $\alpha$ and in (b) we plot
$\lambda\ss{D}(\alpha)=\lambda(t\to\infty,\alpha)$, which increases logarithmically
as $\lambda\ss{D}\approx\ln(1/\alpha)$.
These results are not surprising, even if the actual evaluation of
$\lambda\ss{D}(\alpha)$ should be new at our knowledge. Let us now discuss the main results of
our paper, which are contained in Fig.~\ref{fig_diagram} where we plot
the stability diagram of the Oono-Shiwa equation in the plane $(\alpha,q)$.
The thick line has two branches, corresponding to $q\ss{1,2}(\alpha)$,
the extrema of the interval where the homogeneous solution $u=0$
is linearly unstable. When $\alpha\to\frac{1}{4}$, 
$q\ss{1,2}(\alpha)\to q\ss{c}=1/\sqrt{2}$ and the Eckhaus scenario
applies~\cite{Cross_Greenside}. 
This means that a steady state exists for any $q\in (q_1,q_2)=(q\ss{c}-\delta,
q\ss{c}+\delta)$, where $\delta=(1/4-\alpha)^2$.
If $q\in(q\ss{s1},q\ss{s2})=(q\ss{c}-\frac{\delta}{\sqrt{3}},q\ss{c}+\frac{\delta}{\sqrt{3}})$,
the steady state is stable, otherwise it is unstable and dynamics
proceeds until the final wavevector is in the stable range.
In Fig.~\ref{fig_diagram} we plot as dotted lines the analytical curves
$q\ss{s1,s2}(\alpha)$ in the Eckhaus limit. The range
$10^{-2}< \alpha < \frac{1}{4}$ has been studied in~\cite{Benilov},
we are interested to the regime of small $\alpha$.

Two main questions should be faced. Firstly, how do the extrema of the stable
region, $(q\ss{s1}(\alpha),q\ss{s2}(\alpha))$, vary when $\alpha$ decreases
and more specifically in the limit of vanishing $\alpha$?
We know that for $\alpha=0$ OS equation reduces to the CH equation,
which means that in such limit all the interval $(q_1,q_2)=(0,1)$ is unstable.
If the behavior of the OS equation is continuos in $\alpha=0$,
this limit implies the vanishing of its extrema, $q\ss{s1,s2}(\alpha\to 0)= 0$.
Our results provide an explicit analytical or
numerical evidence for the behavior of $q\ss{s1,s2}(\alpha)$.

The second important question concerns the dynamical evolution of 
the phase separation process: what is the relevant final state?
According to Sec.~\ref{model} 
we know that the free energy ${\cal F}(\alpha,\lambda)$
is minimized for $q\ss{GS}(\alpha)=2\pi/\lambda\ss{GS}(\alpha)\approx
\alpha^{1/3}$. 
However, in the absence of noise the ground state cannot be attained
and even in the presence of noise it might take too long. Therefore,
the knowledge of the ground state is not really helpful for real dynamics.

Figure~\ref{fig_diagram} tries to answer these questions.
Circles and asterisks refer to the final wavelength of some simulations, 
whose initial conditions are of various types: random; a single sinusoidal
profile or a superposition of sinusoidal profiles, with or without noise; 
relaxed asymptotic profiles for slightly different parameters 
(more details are given in~\ref{sec_incon}).
Orange squares simply represent the lower values of asterisks for a given
$\alpha\le 10^{-2}$.
All steady states are limited by a lower and an upper curve, dashed line and dot-dashed line, representing respectively $q\ss{s1}(\alpha)$ and $q\ss{s2}(\alpha)$.
The line $q\ss{s1}(\alpha)\approx \sqrt{\alpha}$ is obtained numerically
by a fit of orange squares
and it vanishes with the same exponent as $q_1(\alpha)\simeq\sqrt{\alpha}$,
see Eq.~(\ref{q12}).
The line $q\ss{s2}(\alpha)$ has been obtained analytically, as we are going
to discuss.

In the Eckhaus limit ($\alpha\to\frac{1}{4}$) the dynamics is described by the
envelope equation, whose stationary solutions are known, see \ref{app_phase}.
Their stability can be analyzed by perturbing the amplitude and the phase.
The amplitude is always stable and it is slaved to the dynamics of the
phase, i.e. the wavevector. The phase dynamics may be stable or unstable.
If we decrease $\alpha$, we move away from the Eckhaus limit and the
exact expression of steady states is not known. However, we can apply
standard analytical approaches~\cite{Whitham,pmPRE1d} and find the formal expression for the
phase diffusion coefficient~\cite{vgoono_pd}:
\be
D = \frac{
q^2\p_q\left( q^{-1}\langle u_x^2-\alpha w^2\rangle\right)}{
\langle u^2\rangle},
\label{eq_D}
\ee
where $u(x)$ is the steady state of period $\lambda=2\pi/q$
and $\p_x w=u$ with $\langle w\rangle=0$.
Here we are assuming that $q\ss{s2}(\alpha)$ is determined by a
long wave instability, which seems to be the case here.

In the coarsening regime, where solutions are unstable, the profile
is similar to what we get for $\alpha=0$, see Fig.~\ref{fig_time}(a).
Therefore, it is reasonable to locate the stability limit $q\ss{s2}(\alpha)$
using the steady states of the CH equation to evaluate the coefficient $D$.
We will do that for large wavelength, where $u(x)$ is a sequence of
remote kinks/antikinks. The single kink profile is $\tilde u(x)=
\tanh (x/\sqrt{2})$. After some easy algebra, similar in spirit to the evaluation
of $\tilde {\cal F}\ss{OS}$, see Eq.~(\ref{Fos}), we find
$\langle u^2\rangle \simeq 1$ and
$ \langle w^2\rangle \simeq \frac{\lambda^2}{48}$.
Along with the evaluation of $\langle u_x^2\rangle$ already given in Eq.~(\ref{ux2}),
we can determine $D$,
\be
D(q)= \frac{\alpha}{16}\lambda^2 - 8 \exp(-\lambda/\sqrt{2}).
\ee
So $D$ becomes positive for $\lambda>\lambda\ss{min}$, where
\be
\frac{\alpha}{16}\lambda\ss{min}^2 = 8 \exp(-\lambda\ss{min}/\sqrt{2}).
\label{theo_estimate}
\ee
The curve $q\ss{s2}(\alpha)=2\pi/\lambda\ss{min}(\alpha)$ is plotted as the dot-dashed line
in Fig.~\ref{fig_diagram}.
A similar result, using a different approach, was found by Villain-Guillot in Ref.~\cite{vgoono_co}.

If $\lambda\ss{GS}$ is not dynamically relevant, what is the relevant final state?
Even if, in the absence of thermalization, the answer depends on the initial state,
it is clear that the homogeneous state ($u(x,0)=0$ plus some noise)
{\it is} a specially important initial state. The full empty dots represent the
final wavelength $\lambda\ss{D}$, also plotted in Fig.~\ref{fig_alpha}(b). 
The fit of the data gives a logarithmic behavior, meaning that 
$\lambda\ss{D}(\alpha)$ scales as $\lambda\ss{min}(\alpha)$,
apart higher order logarithmic corrections.

\begin{figure}
\begin{center}
\includegraphics[width=0.8\columnwidth]{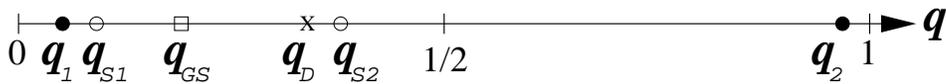}
\end{center}
\caption{
Schematic overview of the different, relevant $q-$values appearing in the paper.
The interval $(q_1,q_2)$ corresponds to linear instability of the homogeneous solution $u=0$.
The interval $(q\ss{s1},q\ss{s2})$ corresponds to linear stability of periodic patterns.
The value $q\ss{GS}$ locates the minimal energy configuration while
$q\ss{D}$ identifies the final state of the deterministic dynamics, starting from the homogeneous solution.
}
\label{fig_q}
\end{figure}

\section{Discussion and conclusions}
\label{discussion}

This work has the main goal to study the dynamical properties of the Oono-Shiwa model
by varying the segregation strength and the initial state of the system.
We can discuss our results by making reference to Fig.~\ref{fig_q}, which summarizes the
various $q-$values which are relevant for the OS equation.

First of all, all displayed $q-$values depend on $\alpha$. The values $q_1$ and $q_2$, explicitely given
in Eq.~(\ref{q12}), represent the extrema of the stability spectrum of the homogeneous solution $u=0$.
When $\alpha\to\frac{1}{4}$ (vanishing segregation), they both tend to $q_c=\frac{1}{2}$ and the system
undergoes the Eckhaus instability~\cite{Benilov}. When $\alpha\to 0$ (strong segregation), $q_1\to 0$ and $q_2\to 1$.
When $\alpha=0$, the OS equation reduces to the CH equation, so all modes $|q|<1$ are unstable and the
system coarsens forever. However, when $\alpha$ is (arbitrarily) small, but finite,
dynamics stops at some finite lengthscale $\lambda\ss{D}$, which depends on the initial state.

The analysis of the minimum of the free energy suggests that $q\ss{GS}=2\pi/\lambda\ss{GS}\simeq \alpha^{1/3}$,
as already found numerically~\cite{Liu_Goldenfeld} and with a similar analytical analysis~\cite{vgoono_en}.
However, the ground state can be attained only if noise is present and the system has enough time to
escape metastable states. In fact, a straight direct simulation of the deterministic dynamics starting from the
homogeneous solution (corresponding to a quenching from the high temperature equilibrium state)
indicates the system coarsens logarithmically for a time $t\ss{D}$ up to $\lambda\ss{D}=2\pi/q\ss{D}
\approx \ln(1/\alpha)$. This scenario can be understood in semiquantitative terms by assuming the term $-\alpha u$
comes into play and stops coarsening when $t\ss{D}\approx\alpha^{-1}$. For smaller time it has negligible effect,
so, if $\lambda\approx \ln t$ is the well known coarsening law for the CH model, it is clear it should be
$\lambda\ss{D} \approx \ln t\ss{D} \approx -\ln\alpha$. This type of reasoning, which is supported
by our numerical results shown in Figs.~\ref{fig_time},\ref{fig_alpha}, is also shown to be correct for
similar models in higher dimension~\cite{Glotzer}.

Therefore, the wavelength of the ground state diverges as $\alpha^{-1/3}$, but coarsening actually stops
way before, at a length $\lambda\ss{D}$ which diverges only logarithmically with $1/\alpha$.
This is due to the fact that the Oono-Shiwa model allows for several (meta)stable states. 
When $\alpha\to\frac{1}{4}$ the Eckhaus scenario applies and the small interval $(q_1,q_2)$, which is
symmetric with respect to $q_c=\frac{1}{2}$, splits into a central interval $(q\ss{s1},q\ss{s2})$
where steady states are stable and two lateral intervals where steady states are unstable for phase
fluctuations. When $\alpha$ decreases and vanishes, also $q\ss{s1}(\alpha)$ and $q\ss{s2}(\alpha)$ vanish.
We suggest that $q\ss{s1}\approx \sqrt{\alpha}$, so that $q_1(\alpha)/q\ss{s1}(\alpha)\to \mbox{const}$.
Instead, $q\ss{s2}(\alpha)$ must vanish slowly, because it must be $q\ss{s2} > q\ss{D}(\alpha)\approx
\ln(1/\alpha)$.

The limits of the stable region $(q\ss{s1},q\ss{s2})$ have been determined either
numerically ($q\ss{s1}$) or analytically ($q\ss{s2}$, although in an approximate way).
We should add that the boundary between unstable steady states ($q>q\ss{s2}$) and metastable steady
states ($q<q\ss{s2}$) may be fairly weak, therefore hard to be detected numerically
(see also Ref.~\cite{Benilov}).
It is interesting that these difficulties do not appear close to $q\ss{s1}$.
As for the nature of the instability occurring at $q\ss{s1,s2}$, 
our analysis suggests that in $q\ss{s2}$ it should be a long-wave instability for any $\alpha$. 
The nature of the instability is less clear in $q\ss{s1}$ and it would require further analysis. 

Finally, we think two other questions have yet to be clarified. 
Firstly, the location of the stable interval $(q\ss{s1},q\ss{s2})$ might be determined
more rigorously with a combined numerical-analytical approach, following Ref.~\cite{Nicoli}.
It comes to having a precise numerical expression of periodic steady states, then
applying Eq.~(\ref{eq_D}) to determine its stability.
Secondly, it would be interesting to add noise to dynamics in order to analyze how
$\lambda\ss{D}$ and $\lambda\ss{GS}$ can reconcile.

We conclude by mentioning the renewed interest towards the problem of phase separation in diblock copolymers,
which may be the way to design new soft materials via self-organization~\cite{PhysicsToday,spatz2000, lopes2001}.
Despite the limitations of the 1D OS model, we think our findings about the role
of dynamics in determining the final pattern are well more general.

\ack
We acknowledge useful discussions with M. B\"ar and S. Alonso
in the first stage of development of the present work, as well
as useful exchange of information with S. Villain-Guillot.
A special thank goes to M. Nicoli, whose collaboration with PP on a 
related problem has been of great help.
AT has received partial financial support from
the German Science Foundation DFG within the framework of SFB 
910 ``Control of self-organizing nonlinear systems`` and
from the Italian MIUR project CRISIS LAB PNR 2011-2013.
PP thanks the Galileo Galilei Institute for Theoretical Physics
for hospitality during the Workshop on Advances in Nonequilibrium
Statistical Mechanics.

\appendix

\section{The Eckhaus instability in the limit $\alpha\to\frac{1}{4}$}
\label{app_Eckhaus}

\subsection{Amplitude stability}

Let us rewrite the Oono-Shiwa equation as follows,
\be
\frac{\p u}{\p t} = -\left( \alpha +\p_{xx} + \p_{xxxx}\right)u + \p_{xx}u^3
\equiv {\cal L}[u] + \p_{xx}u^3
\label{oono2}
\ee
and assume that $\alpha=\frac{1}{4}-\epsilon^2$. 
With this notation, 
\be
{\cal L} = -\left( \frac{1}{2} +\p_{xx}\right)^2 + \epsilon^2
\ee
which is acknowledged to be the linear part of the Swift-Hohenberg (SH) equation. The nonlinear term is different, instead: 
here we have the conserved term $\p_{xx}u^3$, while the term $-u^3$ appears in the SH equation.

Let us now perform a weakly nonlinear analysis when $\epsilon\ll 1$, so 
$\omega(q)=\epsilon^2 -\left( \frac{1}{2} -q^2\right)^2$ and only the interval $(q_c-\delta,q_c+\delta)$
is linearly unstable, with $q_c=1/\sqrt{2}$ and $\delta=\epsilon/\sqrt{2}$.
The solution of the linear part is $u(x,t)=a_q(t)\cos(qx)$, with $a_q(t)=a_q(0)\exp(\omega(q)t)$.
It is customary to expand the nonlinear term in harmonics of the base state and analyze the effect of
the first correction term.

Since $\cos^3 z = \frac{3}{4}\cos z +\frac{1}{4}\cos(3z)$, we get
\be
\dot a_q = \omega(q) a_q - \frac{3}{4} q^2 a_q^3,
\ee
where the troncation is justified because $\omega(3q)<0$, therefore stable~\cite{Cross_Greenside}.
The steady solution
\be
a_q \simeq \frac{2}{q_c}\sqrt{\frac{\omega(q)}{3}}
\ee
is stable and indicates that a steady profile exists in all the region where $\omega(q)>0$.

The statement the profile is stable should be taken with caution, because above calculation means it is
stable against amplitude fluctuations. In order to study phase fluctuations we must go beyond
the single harmonic approximation, which is done with a multiscale analysis. Here below
we perform one such analysis, valid in the limit $\epsilon\to 0$. In \ref{app_phase} we perform
a more general analysis, valid arbitrarily far from the threshold. 

\subsection{Phase stability}
The calculation follows the same lines as for the SH equation, which is well known and discussed in
several books~\cite{Collet_Eckmann,Hoyle,Cross_Greenside}.

We start using the new variables and expansions
\bea
x_0=x, X=\epsilon x, T=\epsilon^2 t \\
u=\epsilon u_1 + \epsilon^2 u_2 +\epsilon^3 u_3.
\eea
At the first order, standard analysis allow to get
\be
u_1(x_0,X,T) = A(X,T) e^{iq\ss{c}x_0} + \mbox{c.c.} .
\ee
The second order gives a similar expression and the third order,
through a solvability condition gives
\be
\p_T A(X,T) = 2\p_{XX}A + \left(1-\frac{3}{2}|A|^2\right) A.
\ee

This result proofs that the dynamics of the OS equation close to the instability threshold
is equivalent to the dynamics of the SH equation in the same limit.
Therefore, the interval $(q_c-\delta,q_c+\delta)$ is divided into three subintervals:
(i) the central interval $(q_c-\delta/\sqrt{3},q_c+\delta/\sqrt{3})$ is stable, the others being
unstable under phase perturbation;
(ii) the left interval $(q_c-\delta,q_c-\delta/\sqrt{3})$ undergoes a splitting process of rolls with
decrease of the local wavelength;
(iii) the right interval $(q_c+\delta/\sqrt{3},q_c+\delta)$ undergoes a coalescence process of rolls,
which increases their local wavelength.

\section{The phase diffusion equation}
\label{app_phase}

Just for completeness, we report here also the main lines of the derivation
of the phase diffusion equation.
It is useful to rewrite the Oono-Shiwa equation in more general terms,
\be
\p_t u = -\p_{xx} [ B(u) + u_{xx} ] -\alpha u
\ee
which will be analysed with a multiscale analysis, whose details can be found elsewhere~\cite{pmPRE1d},
\bea
\p_t &=& \epsilon (\p_T\Psi) \p_\phi \\
\p_{xx} &=& q^2\p_{\phi\phi} + \epsilon\Psi_{XX}(2q\p_q +1)\p_\phi\\
u &=& u_0 +\epsilon u_1 ,
\eea
where the small parameter $\epsilon$ does not appear explicitely in the OS equation: 
it measures in a self consistent way the weak dependence of the phase on space and time. 
The details of the calculation can also be found in Ref.~\cite{vgoono_pd}.

The slow phase $\Psi$ is seen to satisfy a diffusion equation,
$\Psi_T = D \Psi_{XX}$, whose diffusion coefficient,
\be
D(q) = \frac{
q^2\p_q\left( q^{-1}\langle u_x^2-\alpha w^2\rangle\right)}{
\langle u^2\rangle},
\ee
is a function of the properties of steady states of wavevector $q$.
A negative sign of $D$ points out to a phase instability. 

\section{Numerics}
\label{sec_numerics}

\subsection{The numerical integration scheme}
\label{app_code}

To perform the numerical integration of the Oono-Shiwa model
(\ref{oono}) in one spatial dimension over a domain of length $L$, 
we considered a discrete spatial grid of resolution
$\Delta x$ and a discrete time evolution with a time step $\Delta t$.
The discretized field can be written as
$u(i,n)$, where the integer indexes $i$ and $n$ denote
the spatial and temporal discrete variable, respectively.
Periodic boundary conditions have been
considered for the field, i.e. $u(i,n)=u(i+I,n)$, 
where $I$ is the number of sites of the grid ($L=(I-1) \Delta x$).

Following~\cite{frauenkron1994,torcini2002} we have numerically
integrated the Oono-Shiwa model by employing a 
time-splitting pseudo spectral code.
Such integration scheme requires to separate 
the equation in a linear and a non linear part as follows
\be
\frac{\p u}{\p t} = {\cal L}(u) + {\cal NL} (u)
\label{oono_LNL}
\ee
where ${\cal L}(u)= -u_{xxxx}-u_{xx}$ and 
${\cal NL}(u)= 3u^2u_{xx}+6uu_x^2-\alpha u$.
 
As usual for time splitting algorithms, we will
separate the integration step in two successive steps:
each involving only the linear or the nonlinear operator.
Therefore, a complete integration time step will consist 
in the following sequence
\be
 u^\prime(x,t + \Delta t) = {\rm e}^{{\cal L} \Delta t} u(x,t)
\ee
and
\be
 u(x,t+\Delta t) = {\rm e}^{{\cal NL} \Delta t} u^\prime(x,t+\Delta t)
\ee
where $u^\prime(x,t)$ is a dummy field employed during the integration.

Let us now consider the first integration, associated
to the linear operator, namely
\be
\partial_t u(x,t) = {\cal L} u(x,t) \quad  .
\label{lin}
\ee
Eq.(\ref{lin}) can be easily solved in the
Fourier space. The equation of motion for the spatial
Fourier transform
of the field ${\tilde u}(p,t)$ will be
\be
\partial_t {\tilde u}(p,t) = (-p^4 + p^2) {\tilde u}
\label{fftlin}
\ee
The time evolution for ${\tilde u}$ is simply given by
\be
{\tilde u}^\prime (p,t+\Delta t) = \exp{[(-p^4+p^2)\Delta t]} {\tilde u} (p,t)
\enskip .
\label{linsol}
\ee
Therefore in order to integrate Eq.(\ref{lin}), firstly
the field should be Fourier transformed in space (${\cal FT}$),
then it should be multiplied by the propagator reported in Eq.(\ref{linsol})
and the outcome of such operation should be inverse-Fourier
transformed (${\cal FT}^{-1}$), namely:
\be
u^\prime (x,t+\tau) = {\cal FT}^{-1} \exp{[(-p^4+p^2) \Delta t]}
{\cal FT} u (x,t) \,.
\label{linsol2}
\ee

The integration of the nonlinear part has been performed
simply by employing a first order Euler scheme.
However, to obtain a better precision the spatial
derivatives appearing in the nonlinear term have been estimated
employing spectral codes.

\subsection{Simulations Details}
\label{sec_incon}

The simulations have been initialized with different initial conditions.
In particular, the results reported in Figs.~\ref{fig_stability},
\ref{fig_time}, and \ref{fig_diagram} (blue circles) have been obtained
by choosing at time zero the value of the field $u(i,0)$ from a flat 
random distribution,symmetric around zero and of amplitude 0.2.

The data reported as green asterisks in Fig.~\ref{fig_diagram},
denoting stable solutions, have been obtained by starting with 
a sinusoidal configuration of wavelength $q$, namely
$$
u(i,0) = a_0 \cos[q(i-1) \Delta x]
$$
where $a_0 = 0.5$ and $\Delta x$ is the spatial resolution.
In some cases we have verified the stability of the solutions by
adding at time $t=0$ some noise to the above initial configuration. 
In other cases we started from a relaxed configuration and 
we slightly stretched the spatial domain to verify if the solutions remain
stable with a different wave vector.
The orange squares in the same figure indicates the minimal stable
$q$ measured by employing such different initial configurations.

We have usually employed for the analysis of the coarsening
arrest starting from random initial conditions (homogeneous state)
$L = \sqrt{2} \pi \times 40$ , with $\Delta t = 0.0025$ and $I=1024$ 
(corresponding to $\Delta x \simeq 0.17$) with 
total integration times of order $10^5$. 
For the numerical study of the stability of periodic solutions
we used $L = \sqrt{2} \pi \times 100$ , with $\Delta t = 0.025$ and $I=1024$ 
(corresponding to $\Delta x \simeq 0.43$)
and we
have reached integration times up to $T \simeq 10^6$.

A few tests have been performed to ensure that the employed numerical accuracy 
was sufficient. In particular, by decreasing the time step down to $\Delta t = 0.0025$
and the spatial resolution down to $\Delta x = 0.1$ we do not observe any relevant
modification of the obtained results.

\section*{References}

\bibliographystyle{iopart-num}

\end{document}